\documentclass[
    ,final            % use final for the camera ready runs
  ]
  {aipproc}

\usepackage{amssymb}
\usepackage{epsfig}

\layoutstyle{6x9}

\newcommand{\Dphi}{\Delta \phi\,{}_{\rm dijet}}
\newcommand{\ptmax}{p_T^{\rm max}}
\newcommand{\mjj}{M_{\rm jj}}

\begin{document}

%\title{Recent Run II QCD Results from D\O\ }

\title{\vskip-1cm Recent Run II QCD Results from D\O\ \\ \vskip2mm
\footnotesize \rm (To be published in the Proceedings of
the 15$^{\rm th}$ Topical Conference on Hadron Collider Physics,\\  HCP 2004,
Michigan State University, East Lansing, MI, June 14-18, 2004)}

\author{Markus Wobisch  \\ 
         \small  for the D\O\ Collaboration}{
  address={Fermilab, P.O. Box 500, Batavia, Illinois 60510}
}

\begin{abstract}
We present recent QCD results from the D\O\ experiment
at the Fermilab Tevatron Collider
in $p\bar{p}$ collisions at $\sqrt{s}=1.96\,{\rm TeV}$.
Results are presented for the inclusive jet and dijet cross sections,
a measurement of dijet azimuthal decorrelations, 
studies of elastic scattering,
and a search for diffractively produced $Z$ bosons.
\end{abstract}

\maketitle

% ***********************************************************
% ***********************************************************
% ***********************************************************

\section{Introduction}
The center-of-mass energy for proton-antiproton ($p\bar{p}$) 
collisions was increased from $\sqrt{s}=1.8\,{\rm TeV}$ to 1.96\,TeV
in Run II of the Tevatron Collider at Fermi National Accelerator Laboratory.
This article presents QCD results measured with the 
D\O\ detector~\cite{run2det},
based on data sets corresponding to integrated luminosities
of $\approx 150\,{\rm pb}^{-1}$.
These are measurements of the inclusive jet and the dijet cross section,
studies of azimuthal decorrelations in dijet production, 
a search for diffractively produced $Z$ bosons, and 
a measurement of the $dN / d|t|$ distribution in elastic $p\bar{p}$
scattering.

% ***********************************************************
% ***********************************************************
% ***********************************************************
\section{Jet Production}

\begin{figure}
  \includegraphics[scale=0.91]{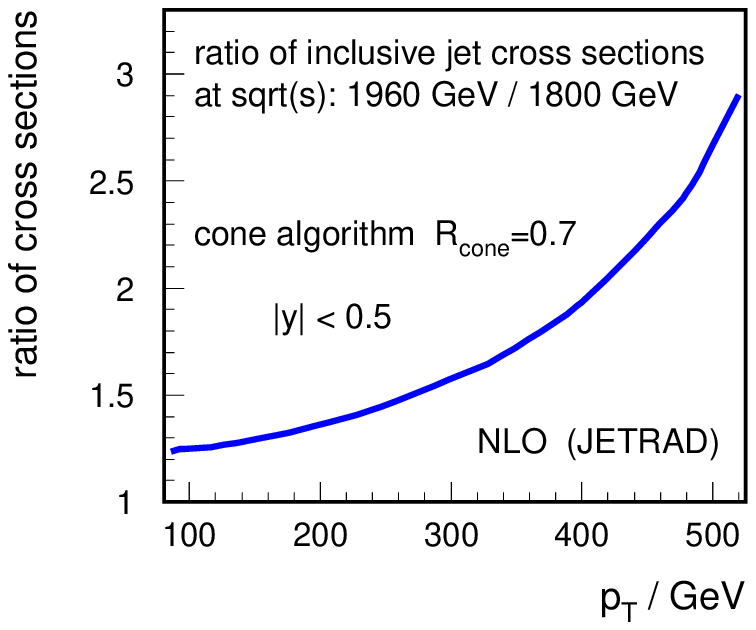}\hskip4mm
  \includegraphics[scale=0.91]{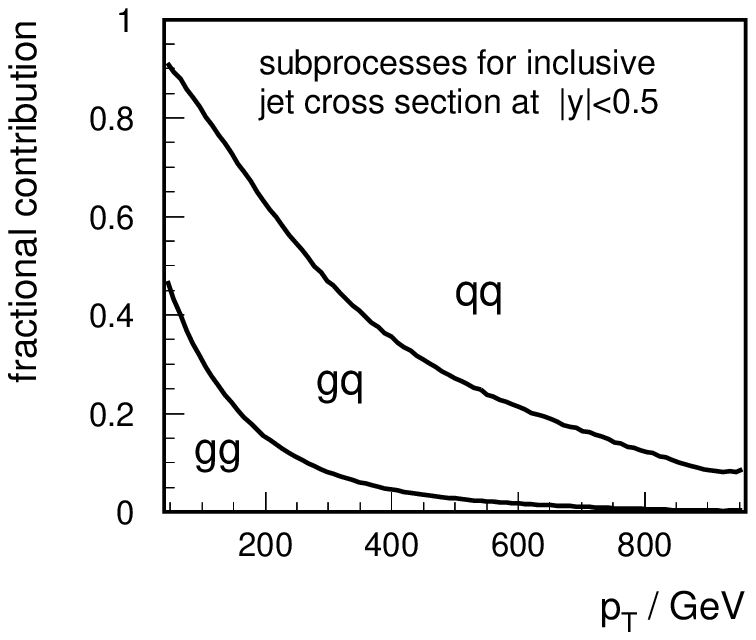}
  \caption{\label{theo}
            The ratio for the NLO pQCD predictions for the 
            central inclusive jet cross section at different
            center-of-mass energies as a function of jet $p_T$ (left).
            The decomposition of the central inclusive jet 
            cross section into different partonic subprocesses
            as a function of jet $p_T$ (right).
}
\end{figure}

The production rates of particle jets with large transverse momentum
($p_T$) provide sensitive tests for the predictions of 
perturbative QCD (pQCD) and give information on the non-perturbative
structure of the proton as parameterized in the
parton density functions (PDFs).
The increased center-of-mass energy of $\sqrt{s}=1.96$\,TeV 
in Run II results in an increased jet cross section, as compared to 
$\sqrt{s}=1.8$\,TeV in Run I.
Fig.~\ref{theo} shows the ratio of the pQCD predictions for 
the inclusive jet cross section for both center-of-mass energies.
At $p_T$ above $400$\,GeV the cross section increases by more than a factor 
of two.
At high integrated luminosity this increases the accessible $p_T$ range 
and extends the discovery potential for new physics phenomena.
The increased cross section at high $p_T$ will also allow to test 
the dynamics of the strong interaction in unprecedented regions
and constrain the PDFs at highest momentum fractions $x$.
The sensitivity of the jet cross section to the PDFs can be
seen from the decomposition of the total jet
cross section into partonic subprocesses.
This is shown in Fig.~\ref{theo} (right) for the  inclusive jet
cross section as a function of $p_T$.
Up to $p_T = 200$\,GeV (where the PDFs are probed at $x\approx0.4$),
the largest fraction of the jet cross section 
comes from gluon-induced subprocesses. 
There are still significant contributions, even at high $p_T$,
from quark-gluon scattering, giving direct sensitivity to 
the gluon density in the proton even at highest $x$.

A major difference between the Tevatron jet measurements 
in Run I and Run II is the jet definition:
the cone algorithm used in Run I was not infrared safe
and reliable pQCD predictions were not possible~\cite{seymour1997}.
The ``Run II cone algorithm''~\cite{run2cone} is still an iterative 
seed-based cone algorithm, 
but it uses midpoints between pairs of proto-jets as additional seeds 
which makes the procedure infrared safe~\cite{seymour1997}.
A further difference between both jet algorithms is the chosen 
``recombination scheme''.
The Run I algorithm used the ``$E_T$-scheme''
(or ``Snowmass convention''~\cite{snowmass})
in which the transverse jet energy $E_T$ was defined as the scalar sum 
of the particle $E_T$'s and the invariant jet mass was defined to be zero.
Consequently, in Run I, the jet kinematics were described by
$E_T$ and pseudorapidity, $\eta = -\ln(\tan(\theta/2))$.
In Run II particles are combined into jets using the ``$E$-scheme''
(addition of four-vectors).
The $E$-scheme results in massive jets which are described by
$p_T$ and rapidity,
$y = \frac{1}{2} \ln{\bf (}(E+p_z)/(E-p_z){\bf )}$.
The $E$-scheme has the advantage that the kinematic boundary
$p_T^{\rm max} =\sqrt{s}/2$ is stable, independent of the number
of particles in a jet (which was not the case in the $E_T$-scheme).
This is desirable for resummed calculations~\cite{leshouches1999}.

All jet measurements presented here are using a data sample 
corresponding to an integrated luminosity of $\approx 150$\,pb$^{-1}$.
The cone radius (in $y$ and $\phi$ space) was chosen to be $R_{\rm cone}=0.7$.
Jet detection was based on a compensating, finely segmented, 
liquid-argon and uranium calorimeter that provided nearly full solid-angle 
coverage.  
Events were acquired using multiple-stage inclusive-jet triggers.  
Analysis regions were defined based on the jet with largest
$p_T$ in an event ($\ptmax$) with the requirement that the trigger
efficiency be at least 99\%.
The position of the $p\bar{p}$ interaction was reconstructed using a
tracking system consisting of silicon
microstrip detectors and scintillating fibers located within a 
$2\,\rm{T}$ solenoidal magnet.
The $p_T$ of each jet was corrected for calorimeter showering effects,
overlaps due to multiple interactions and event pile-up, calorimeter 
noise effects, and the energy response of the calorimeter. 
The data were corrected for selection efficiencies and for
migrations due to resolution.
The largest experimental uncertainty is due to the jet energy 
calibration.

% ***********************************************************
% ***********************************************************
\subsection{Inclusive Jet and Dijet Jet Cross Sections}

\begin{figure}
  \includegraphics[scale=0.39]{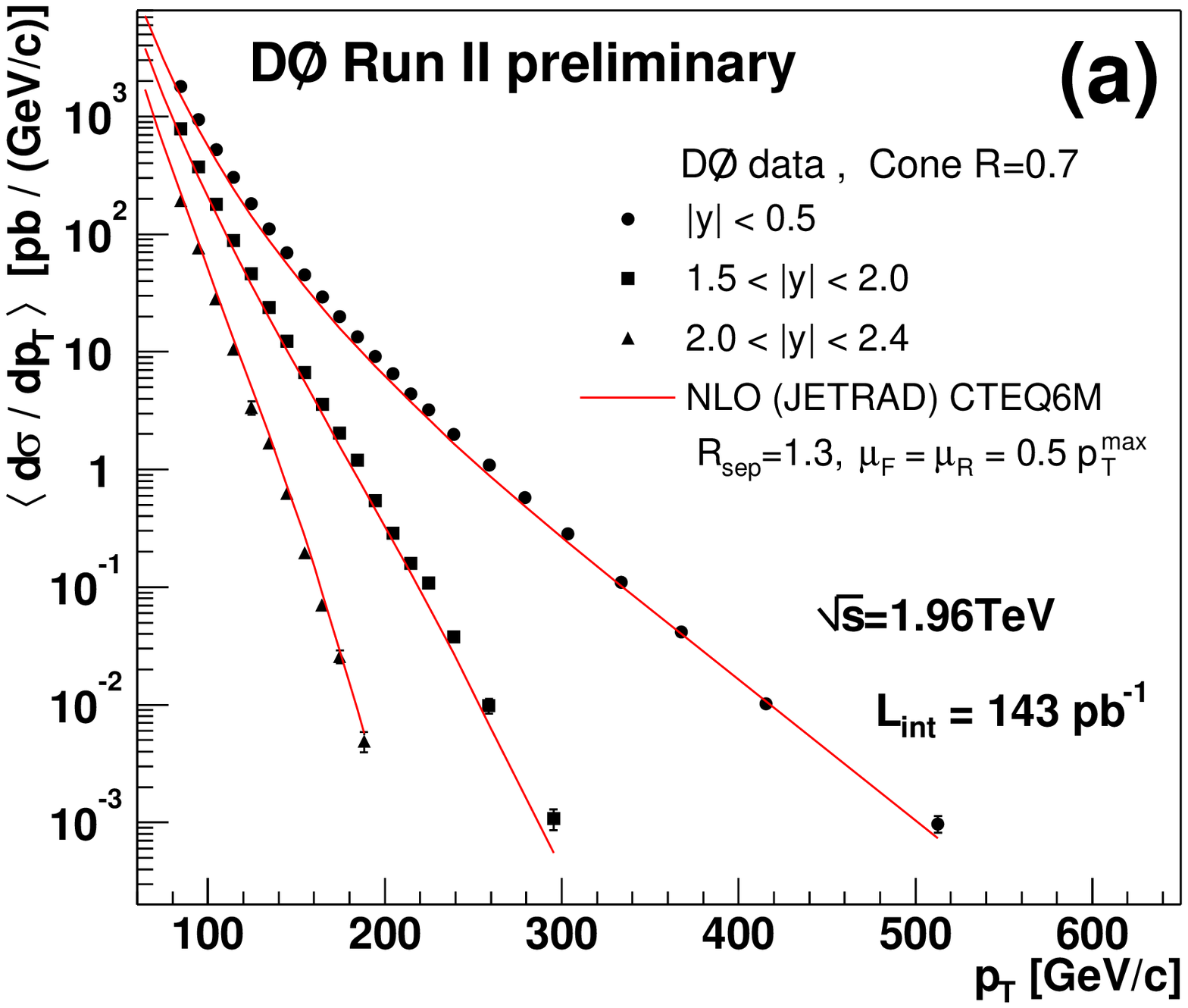} \hskip2mm
  \includegraphics[scale=0.39]{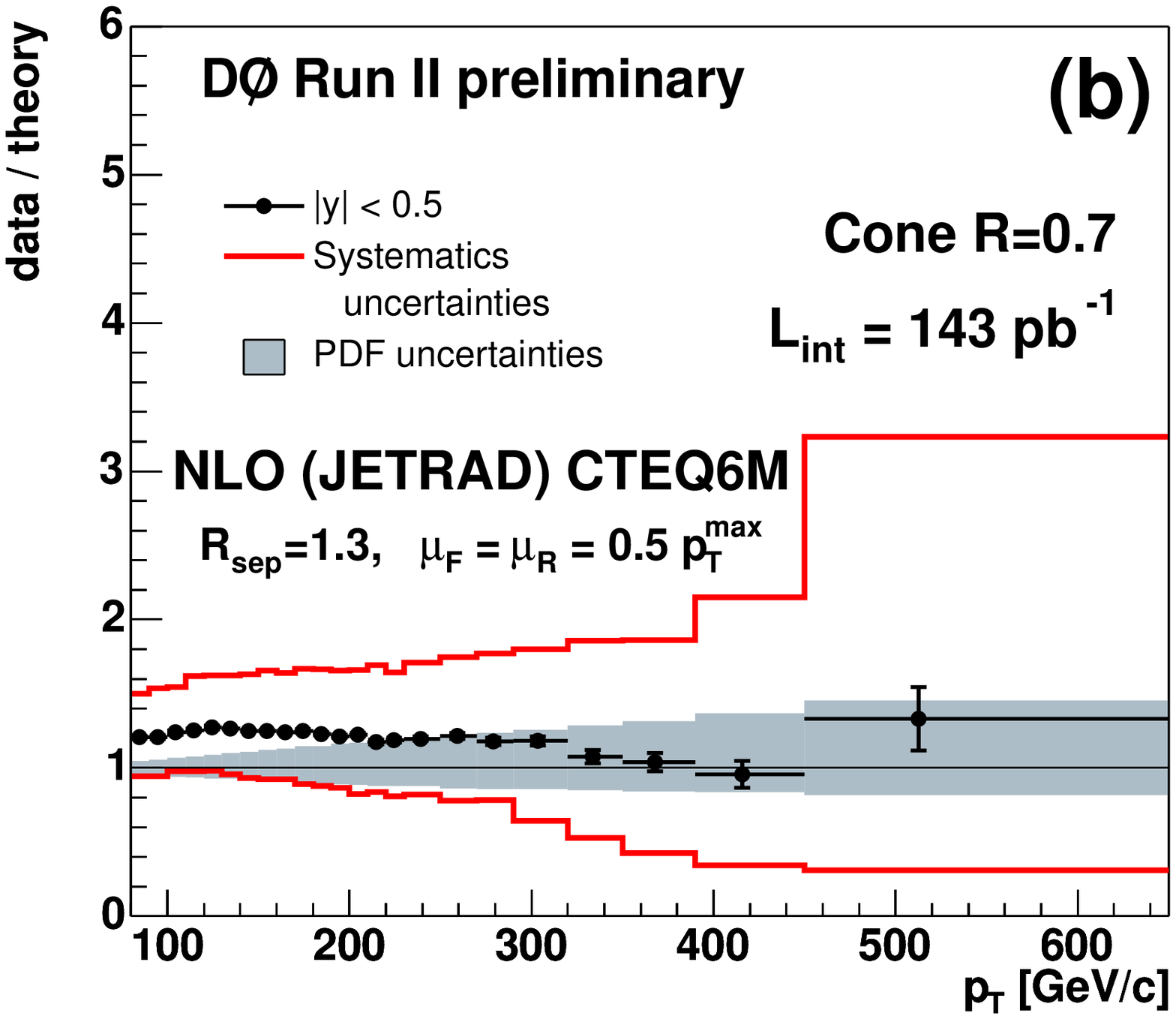}
\end{figure}
\begin{figure}
  \includegraphics[scale=0.39]{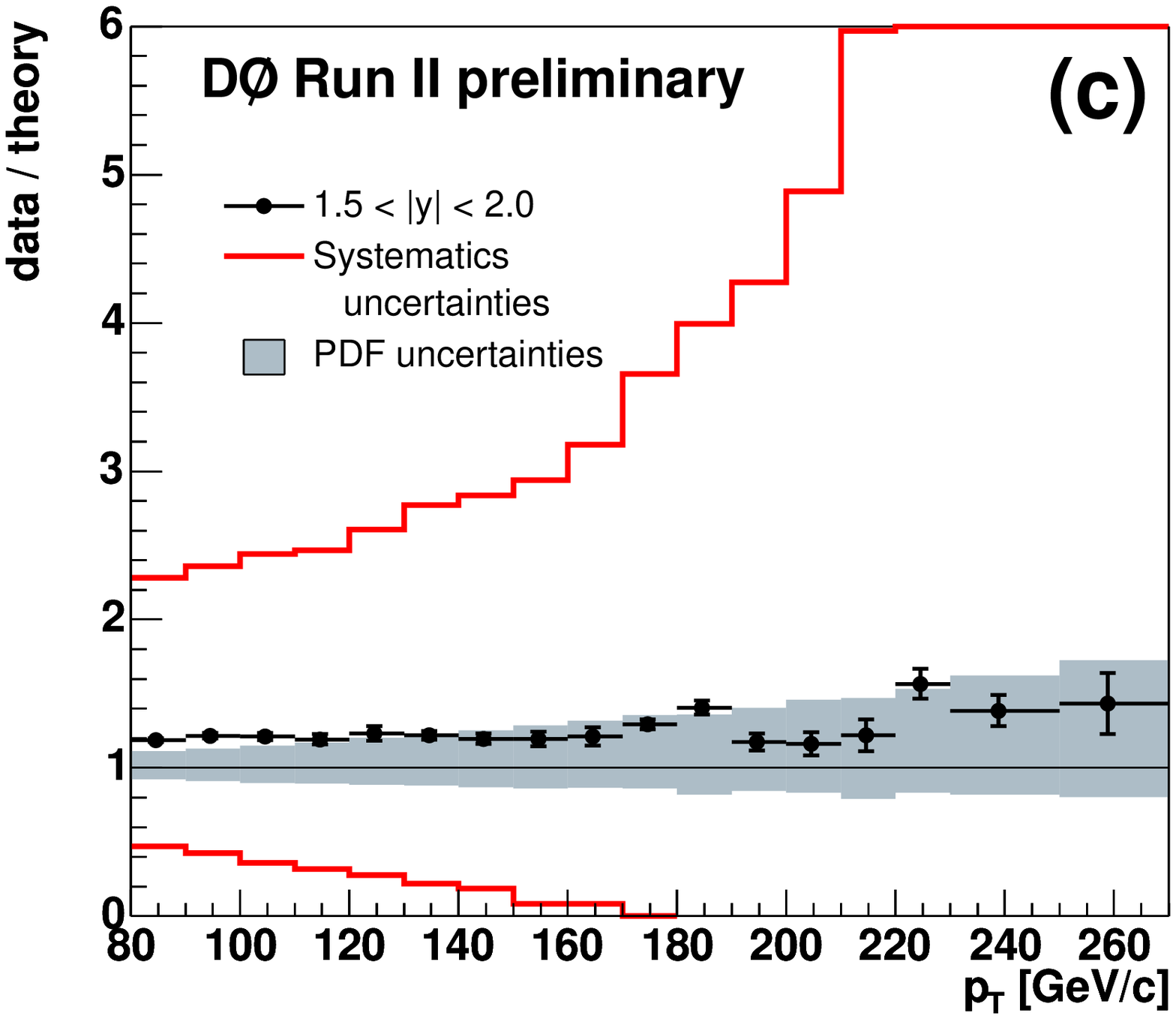} \hskip2mm
  \includegraphics[scale=0.39]{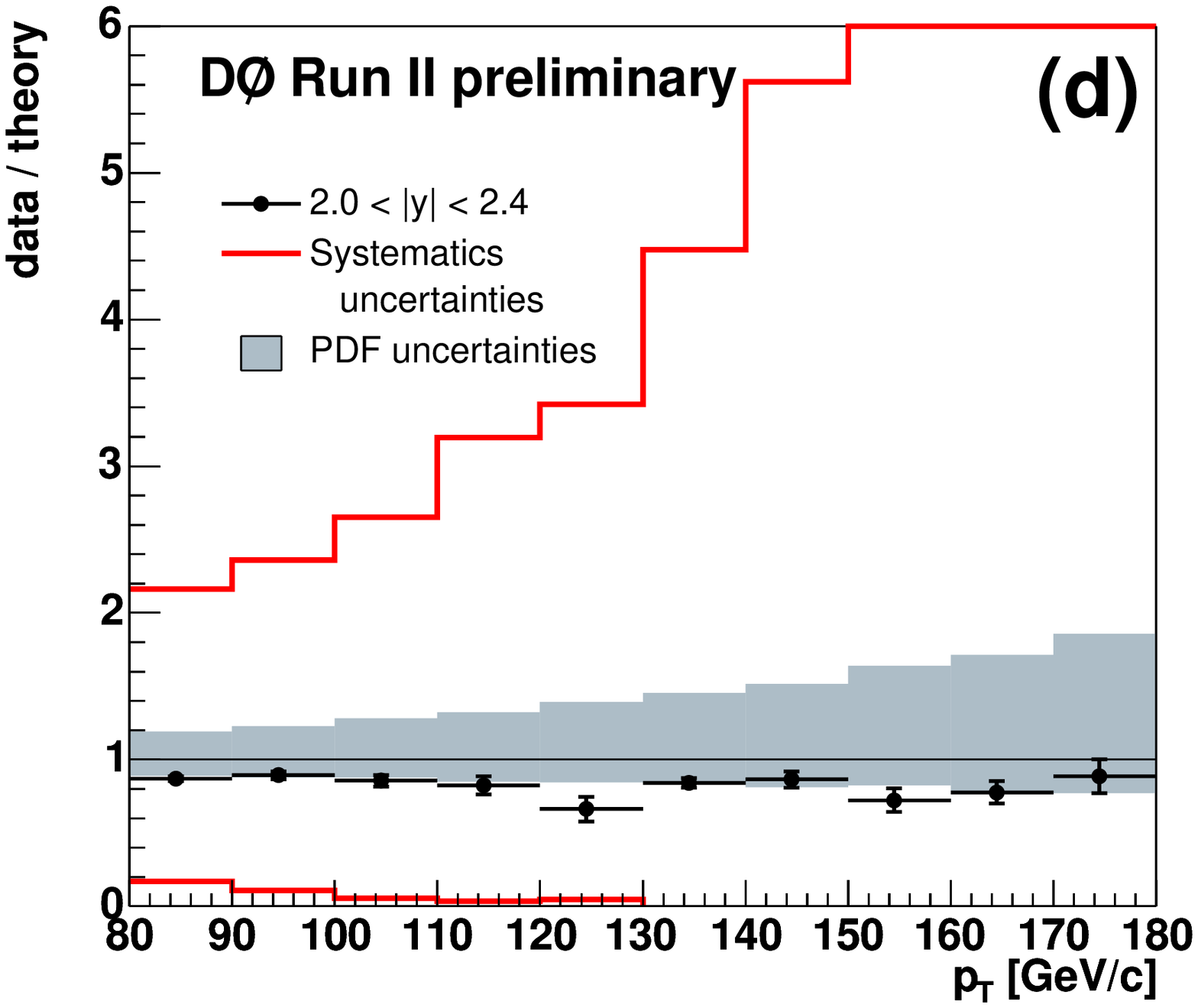}
 \caption{\label{incl1}
        The inclusive jet cross section as a function of $p_T$
        in different rapidity regions (a). Only statistical
        errors are shown. 
        Figures (b), (c), (d) show the ratios of data and the NLO
        pQCD predictions.
        The error bars indicate the statistical errors and the
        total experimental uncertainty is displayed by the lines.
        Theoretical uncertainties due to the PDFs are shown
        as the shaded bands.
 }
\end{figure}

\begin{figure}
  \includegraphics[scale=0.39]{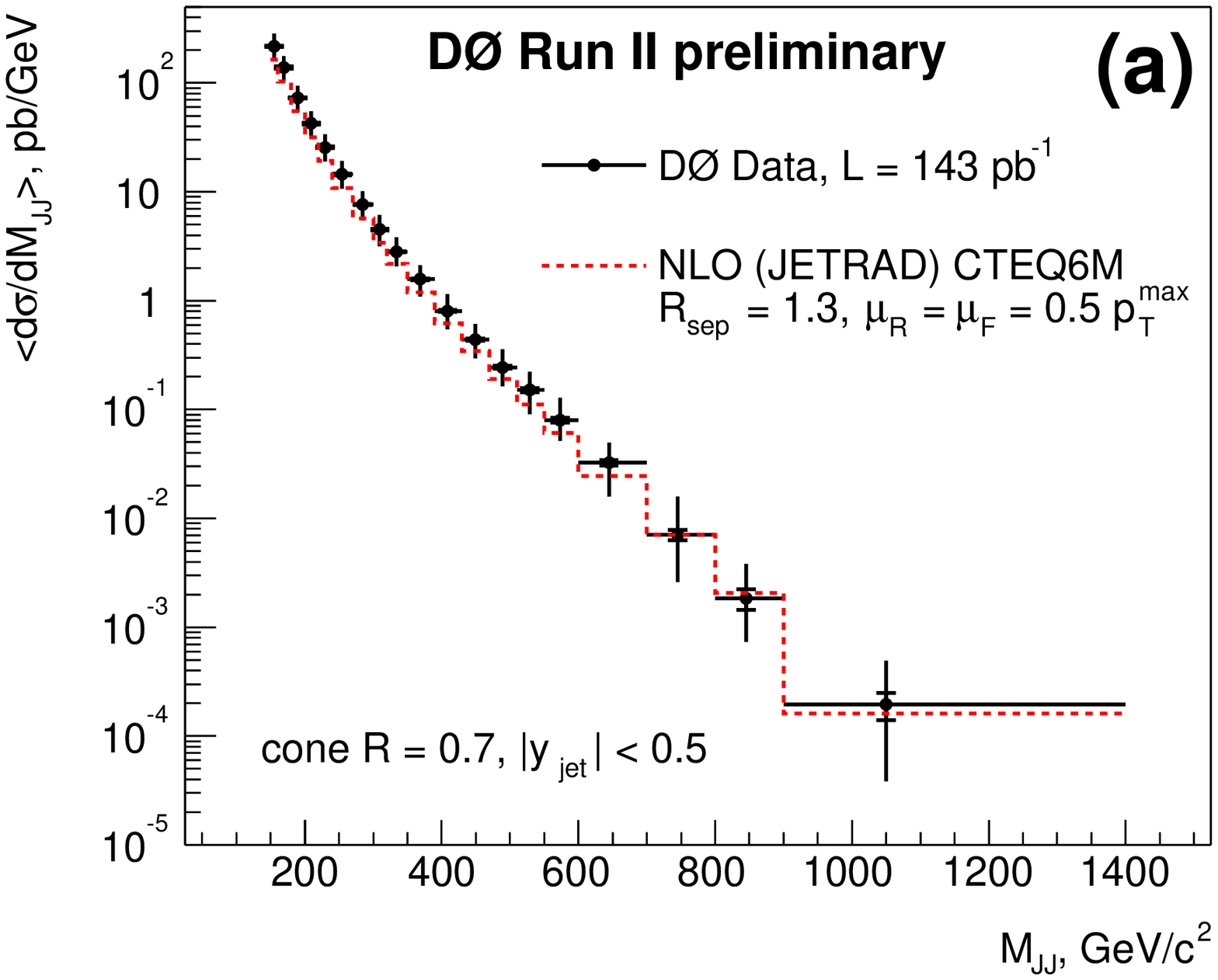}
  \includegraphics[scale=0.39]{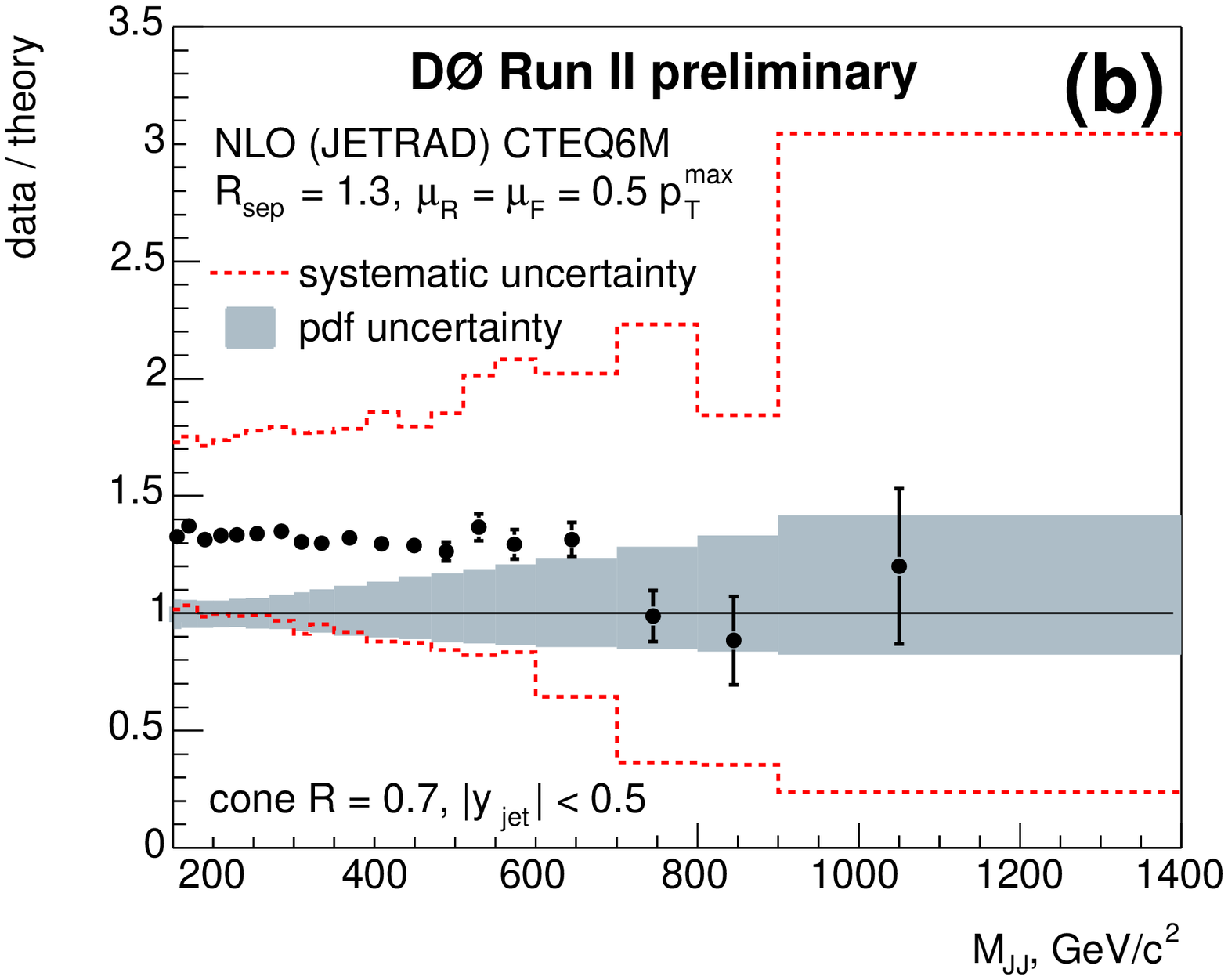}
  \caption{\label{dij1}
     The dijet cross section as a function of $\mjj$ at central
     rapidities. In Fig.~(a) the data are shown with  the total 
     experimental uncertainties. The statistical
     uncertainties are displayed by the inner error bars.
     Fig.~(b) shows the ratio of data and the NLO pQCD
     predictions.
         The error bars indicate the statistical errors and the
        total experimental uncertainty is displayed by the lines.
        Theoretical uncertainties due to the PDFs are shown
        as the shaded bands.
}
\end{figure}

The inclusive jet cross section is measured as a function of $p_T$
and $y$.
Fig.~\ref{incl1} (a) shows the $p_T$ dependence of the inclusive 
jet cross section in three rapidity regions in the range $0 < |y| < 2.4$
(with statistical errors only).
The cross section is falling by more than six orders of magnitude
between $p_T=100$\,GeV and $p_T=500$\,GeV.
At forward rapidities the $p_T$ dependence is much stronger.
Next-to-leading order (NLO) pQCD predictions are overlaid on the data.
The NLO calculations~\cite{jetrad} were computed for
renormalization and factorization scales  $\mu_r=\mu_f= 0.5 \, \ptmax$
using the CTEQ6M~\cite{cteq6} PDFs and $\alpha_s(M_Z)=0.118$.  
The maximum distance of particles within a jet was limited to
$\Delta R <R_{\rm sep} \cdot R_{\rm cone}$ with 
$R_{\rm sep} = 1.3$~\cite{rsep}.
The ratio of data over theory is shown in Figs.~\ref{incl1} (b), (c), (d).
Uncertainties in the NLO calculations due to the
PDFs  are indicated by the grey bands; the experimental
uncertainties are displayed as lines.  
The latter increase with $p_T$, especially at large rapidities.  
Theory has good agreement with data given the large uncertainties.
The dijet cross section at central rapidities
($|y| < 0.5$) is shown in Fig.~\ref{dij1} (a)
as a function of the invariant dijet mass $\mjj$ in the range
$150 < \mjj < 1400$\,GeV.
The ratio of data and theory is displayed in Fig.~\ref{dij1} (b).
Within the large experimental uncertainties theory gives a good 
description of the data.

It is clear that the current jet cross section measurements
with their large uncertainties can not constrain the PDFs.
Improvements in the jet energy calibration and a reduction 
of the corresponding uncertainty are still needed. 
It has to be noted that at the time this article is written 
significant improvements have already been achieved and further
work is in progress.

% ***********************************************************
% ***********************************************************
\subsection{Dijet Azimuthal Decorrelations}

As long as the uncertainties of the jet energy calibration are still
large it is convenient to investigate observables which have 
a smaller dependence on the jet energy calibration but which are still 
sensitive to QCD effects.
The decorrelation of the azimuthal angles of the two leading 
$p_T$ jets, $\Dphi = |\phi_{\rm jet1} - \phi_{\rm jet2}|$,
is such an observable.
In lowest order pQCD both jets in a dijet event have equal
$p_T$ and correlated azimuthal angles $\Dphi=\pi$.
Additional soft radiation causes small azimuthal decorrelations,
whereas $\Dphi$ significantly lower than $\pi$ is evidence of additional hard 
radiation with high $p_T$.
Exclusive three-jet production populates $2\pi/3 < \Dphi < \pi$
while smaller values of $\Dphi$ require additional radiation such as a 
fourth jet in an event.
Distributions in $\Dphi$ provide an ideal testing ground for 
higher-order pQCD predictions without requiring the reconstruction
of additional jets
and offer a way to examine the 
transition between soft and hard QCD processes based on a 
single observable.  

\begin{figure}
  \includegraphics[scale=0.85]{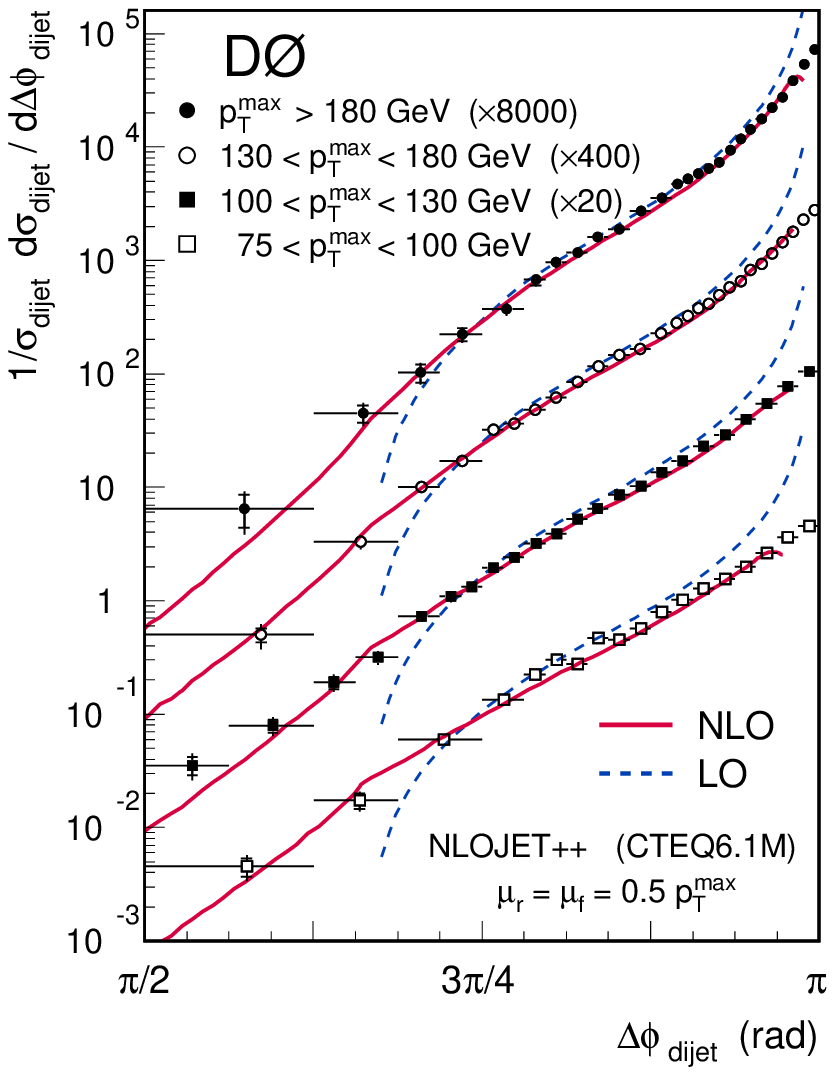}
  \includegraphics[scale=0.85]{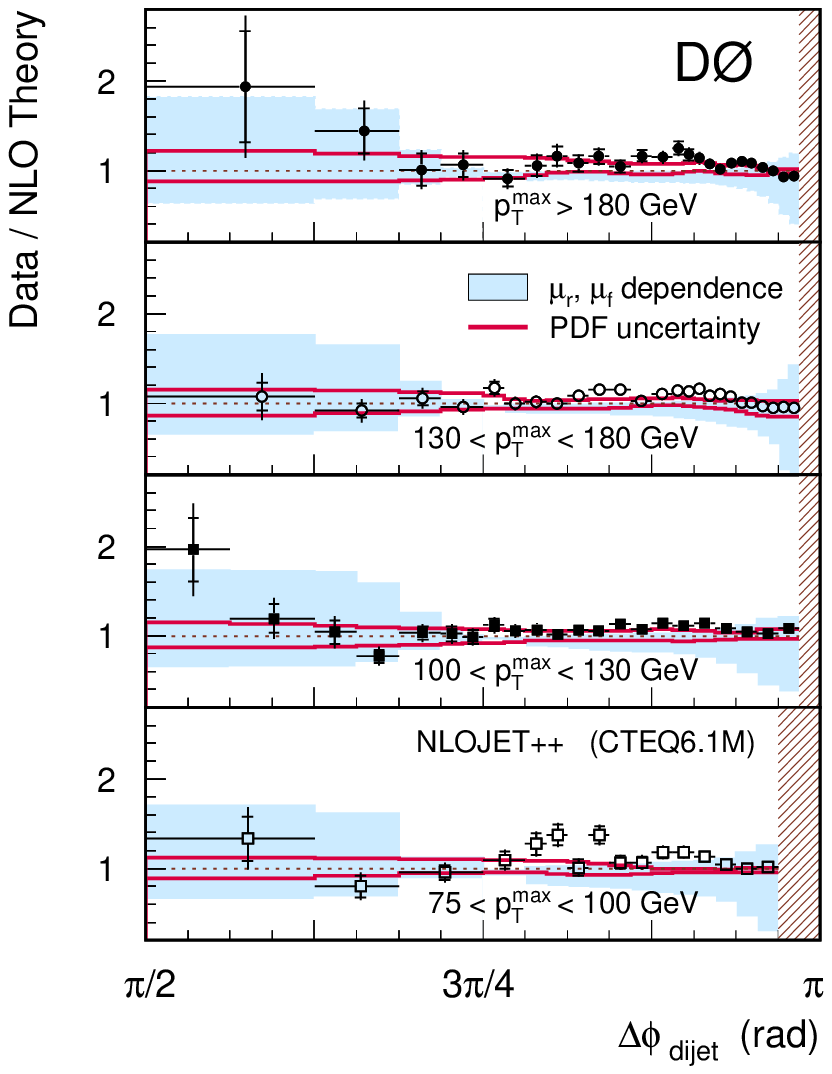}
  \caption{\label{dphi1}
       Left: The $\Dphi$ distributions in four regions of $\ptmax$.  
       The solid (dashed) lines show the NLO (LO) pQCD predictions.
       Right: Ratios of data to the NLO pQCD calculation
             for different regions of $\ptmax$.  
          Theoretical uncertainties due to variation of
          $\mu_r$ and $\mu_f$ are shown as the shaded regions;
          the uncertainty due to the PDFs is indicated by the solid 
          lines. 
          The points at large $\Dphi$ are excluded because the 
          calculation has 
          non-physical behavior near the divergence at $\pi$.
}

\end{figure}

The observable is defined as the differential dijet cross section
in $\Dphi$, normalized by the inclusive dijet cross section 
integrated over the same phase space
$(1/\sigma_{\rm{dijet}})\, (d\sigma_{\rm{dijet}}/ d\Dphi)$.
The corrected data~\cite{dphi} are presented in Fig.~\ref{dphi1} (left) 
as a function of $\Dphi$ in four ranges of $\ptmax$. 
The inner error bars represent the statistical uncertainties and the 
outer error bars correspond to the total experimental uncertainties.
The spectra are strongly peaked at $\Dphi\approx\pi$ and
the peaks are narrower at larger values of $\ptmax$.
The data in Fig.~\ref{dphi1} (left) are compared
to the pQCD predictions in leading order (LO) and NLO~\cite{nlojet} for
$\mu_r=\mu_f=0.5\,\ptmax$ using  CTEQ6.1M~\cite{cteq6} PDFs.
The (N)LO predictions 
are computed as the ratio of the (N)LO predictions 
for $2\rightarrow 3$ processes and  $2\rightarrow 2$ processes.
Due to kinematic constraints, the LO calculation is restricted
to $\Dphi > 2\pi/3$.
At $\Dphi = \pi$ where the third jet becomes soft the
LO predictions diverge.
NLO pQCD provides a good description of the data.  
This is emphasized
in Fig.~\ref{dphi1} (right) which displays the ratio of data and NLO.  
There is good agreement  within a small offset of 
5--10\% relative to unity.  
Theoretical uncertainties due to the PDFs~\cite{cteq6} are estimated
to be below 20\%.
Also shown are the effects of $\mu_r$ and $\mu_f$ variations
in the range $0.25\,\ptmax < \mu_{r,f} < \ptmax$.  
The large scale dependence for $\Dphi<2\pi/3$ occurs because the 
NLO calculation only receives contributions from tree-level four-parton 
final states in this regime.
NLO pQCD fails to describe the data in the region  $\Dphi\approx\pi$ which is
dominated by soft processes and therefore can not be described in
fixed-order perturbation theory.  
pQCD results at large $\Dphi$ in
Figs.~\ref{dphi1}  were excluded because the
calculations have non-physical behavior near the divergence at $\pi$
and were not stable.

\begin{figure}
  \includegraphics[scale=0.85]{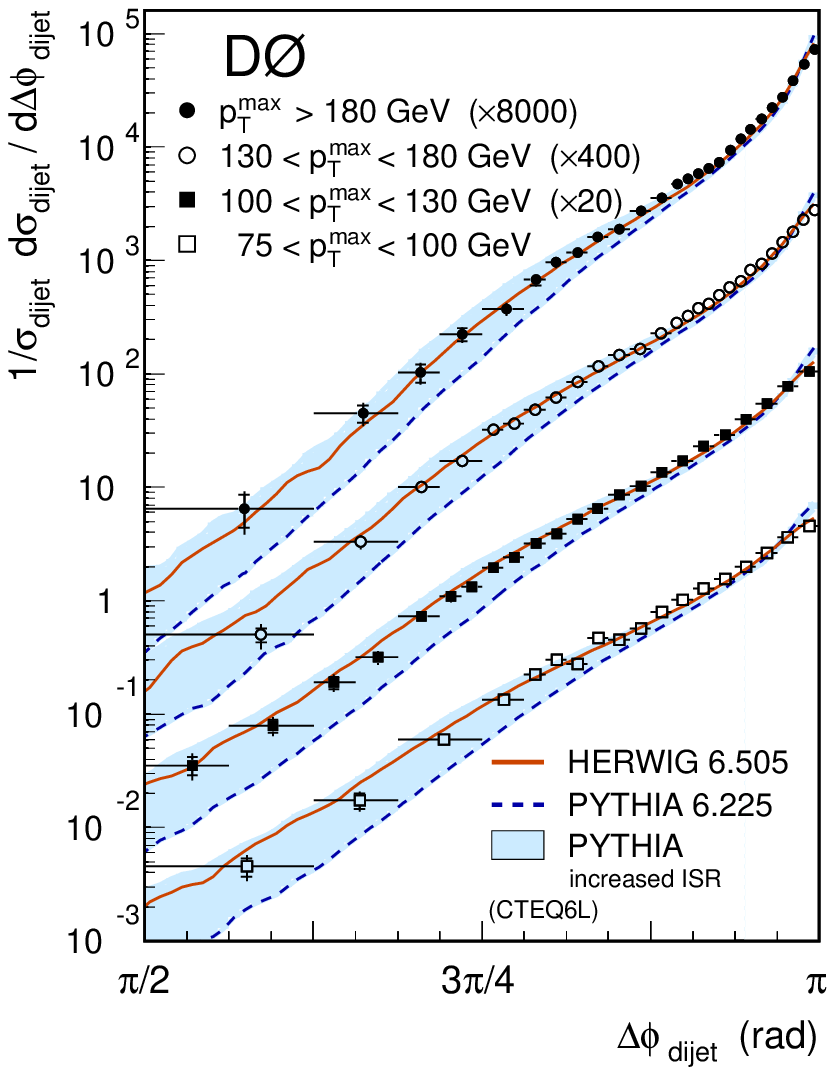}
  \includegraphics[scale=0.85]{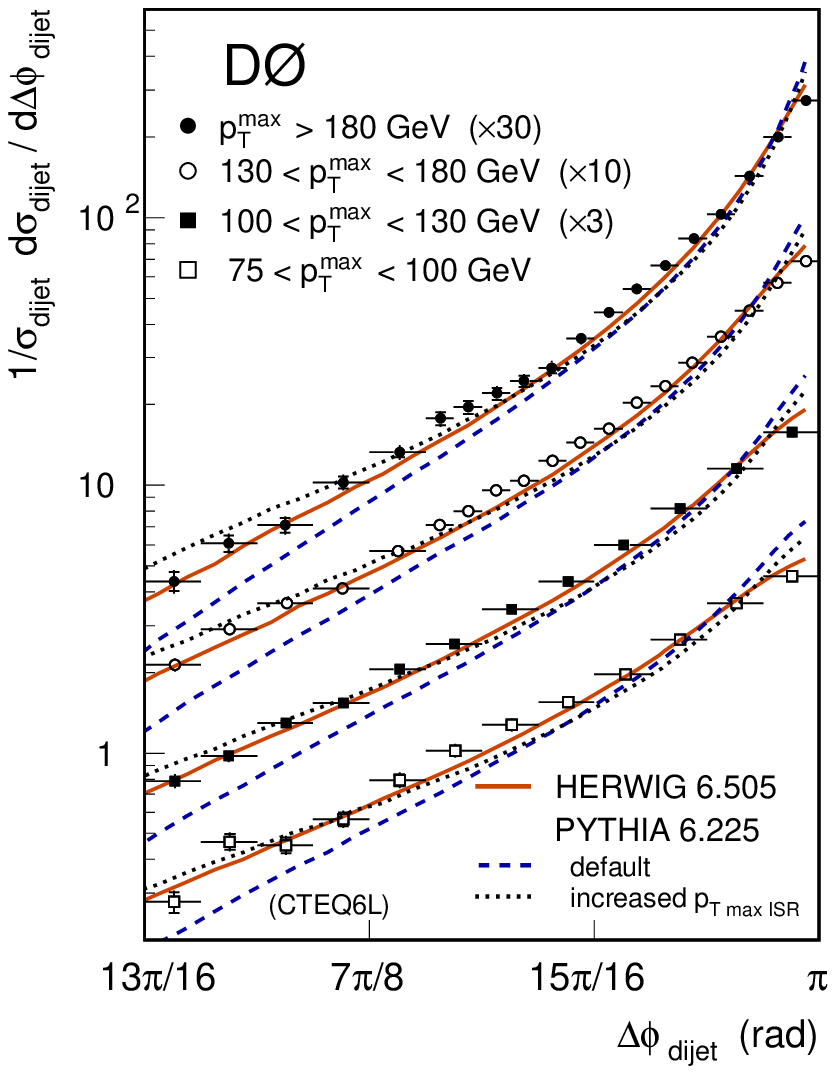}
  \caption{\label{dphi2}
     The $\Dphi$ distributions in different $\ptmax$ ranges.  
     Results from {\sc herwig} and {\sc pythia} are overlaid on the data.
     The left Figure displays the full $\Dphi$ range and 
     the right Figure emphasizes the region close to $\pi$.
}
\end{figure}

Monte Carlo event generators, such as {\sc herwig} and {\sc pythia},
use $2 \rightarrow 2$  LO pQCD matrix elements with phenomenological 
parton-shower models to simulate higher order QCD effects.  
{\sc herwig} and {\sc pythia} results for default parameters, 
CTEQ6L~\cite{cteq6} PDFs, and $\alpha_s(M_Z)=0.118$, 
are compared to the data in Fig.~\ref{dphi2} over the whole 
$\Dphi$ range (left) and in greater detail in the region 
$\Dphi > 13\pi/16$ (right).
The default version of {\sc herwig} gives a good
description of the data over the whole $\Dphi$ range in all $\ptmax$ regions.
It is slightly below the data around $\Dphi \approx 7\pi/8$
and slightly narrower peaked at $\pi$.
The default version of {\sc pythia} does not describe the data.
The distribution is too narrowly peaked at $\Dphi\approx\pi$ and lies
significantly below the data over most of the $\Dphi$ range
in all $\ptmax$ regions.

To investigate the possibilities of tuning  {\sc pythia}
we focus on the impact of the initial-state radiation (ISR)
parton shower and the parameters by which the ISR shower
can be adjusted.
The maximum allowed $p_T$ in the ISR shower is
controlled by the product of the hard scattering scale ($p_T^2$) 
and the parameter \texttt{PARP(67)}.  
Increasing this cut-off from the current default of
$1.0$ to the earlier default of $4.0$ 
leads to significant changes of the {\sc pythia} prediction for $\Dphi$
(the shaded bands in Fig.~\ref{dphi2}, left, and the dotted line
in Fig.~\ref{dphi2}, right).
At low $\Dphi$ the description of the data becomes very good
but at large $\Dphi$ this parameter has not enough effect
to bring {\sc pythia} close to the data.
The best description of the low $\Dphi$ region is obtained
for \texttt{PARP(67)=2.5} (not shown).

\begin{figure}
 \includegraphics[scale=0.85]{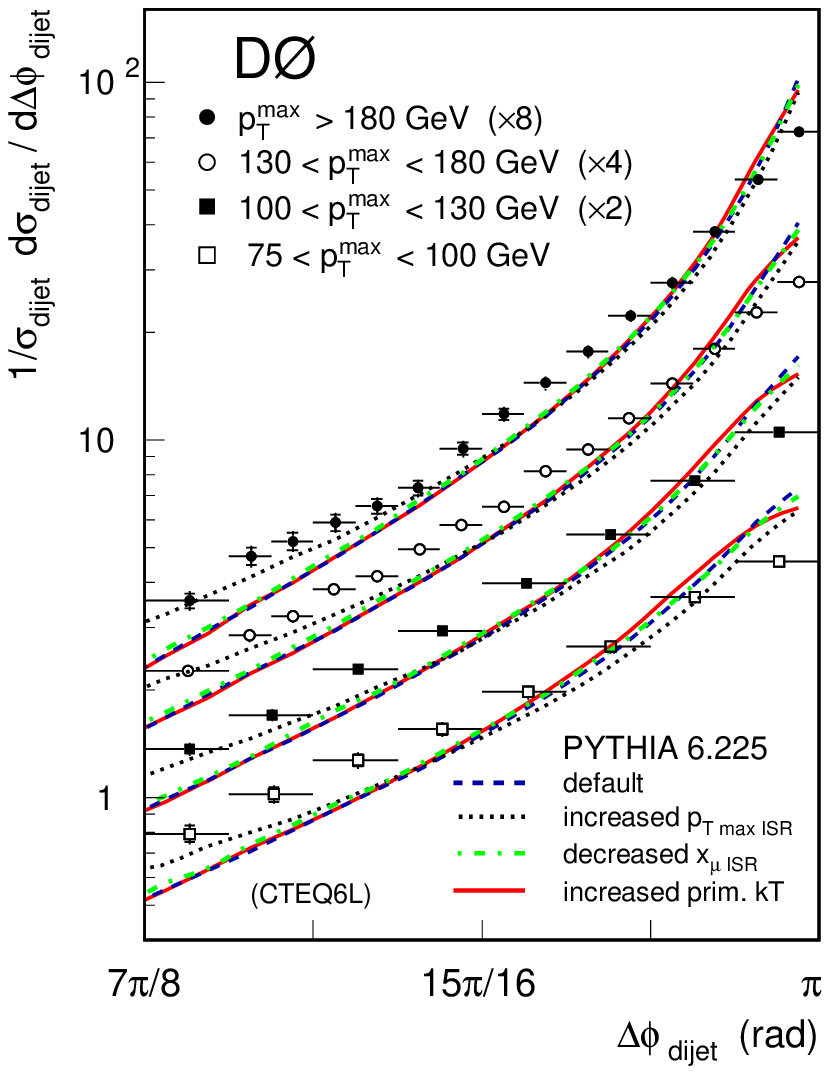}
  \includegraphics[scale=0.85]{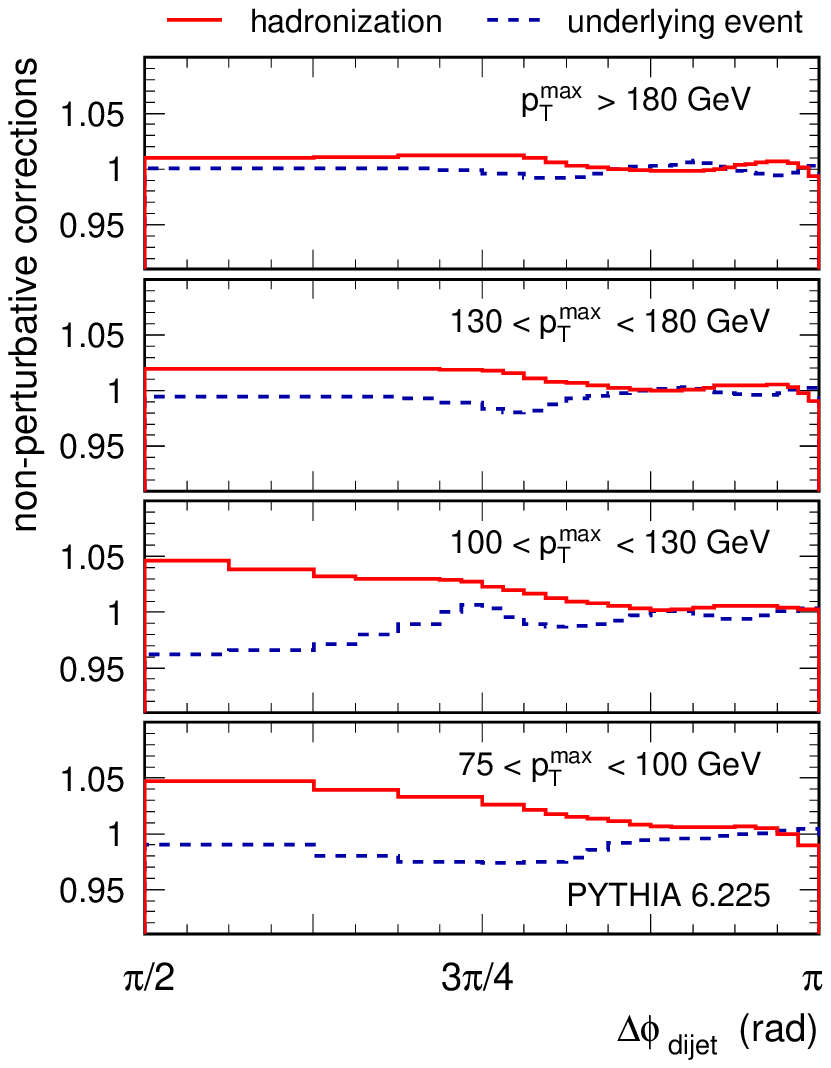}
  \caption{\label{dphi3}
       Left:
      The $\Dphi$ distributions in different $\ptmax$ ranges.  
     Results from {\sc pythia} are overlaid on the data
     for different parameter settings of the ISR shower.
       Right: Non-perturbative corrections for the $\Dphi$
       distribution. Shown are the hadronization corrections
       (solid line) and the effects from underlying event
       as determined using {\sc pythia}.
  }
\end{figure}

Further ISR-related parameters are tested if they have impact on the
$\Dphi$ distribution. 
These are the scale factor, $x_\mu$, for the renormalization scale for 
$\alpha_s$ in the ISR shower, {\tt PARP(64)}, and the primordial 
$k_T$ of partons in the proton: 
the central value of the gaussian distribution,
{\tt PARP(91)}, and the upper cut-off, {\tt PARP(93)}.
We have lowered the factor for the renormalization scale
to {\tt PARP(64)=0.5 (D=1.0)} which increases the value of $\alpha_s$
according to the RGE.
We have alternatively increased the primordial $k_T$ from 1\,GeV
to 4\,GeV, {\tt PARP(91)=4.0 (D=1.0)}, and the upper cut-off
of the gaussian distribution from 5\,GeV to 8\,GeV,
{\tt PARP(93)=8.0 (D=5.0)}.
These parameter variations have no effect on the region
at low $\Dphi$ and the effects at large $\Dphi$ are shown in 
Fig.~\ref{dphi3} (left).
Both effects are very small:
while the scale factor has almost zero influence (dashed-dotted line)
there is some small change for the increased primordial $k_T$
(solid line) which manifests itself, however, only very close to the 
peak region and only at lower $\ptmax$.

Non-perturbative corrections for the $\Dphi$ distribution
are investigated in Fig.~\ref{dphi3} (right) using {\sc pythia}.
Hadronization corrections are defined as the ratio 
of the observable, defined on the level of stable particles
and on the level of partons. 
The underlying event correction
is defined as the ratio of the predictions
including underlying event and with the
underlying event switched off.
Both corrections are below 5\%, decreasing with $\ptmax$.
The smallness of these corrections allows to directly compare
the measurement to purely perturbative QCD predictions and it also allows
to interpret the discrepancies of the Monte Carlo event 
generators in terms of perturbative processes.

% ******************************************************************
% ******************************************************************
% ******************************************************************
\section{Elastic and Diffractive Scattering}

So far we have discussed inelastic processes with 
large transverse momentum. 
However, only a small contribution to the total $p{\bar p}$ cross section
comes from such hard inelastic processes.
About 40\% of the total $p{\bar p}$ cross section is
elastic scattering and diffraction.
In elastic scattering both protons are scattered with no momentum loss
under a small angle and emerge intact.
No further particles are produced.
In single diffraction only one of the protons remains intact and
the other proton dissociates.
While additional particles can be produced,
there is often a large rapidity interval towards the intact 
proton in which no particles are emitted (rapidity gap).
Both elastic scattering and diffractive processes
can be identified experimentally by either measuring the
scattered protons or by reconstructing the rapidity gap 
as an angular region with very little energy in the calorimeter.
In the following sections we present a search for diffractively produced 
$Z$ bosons in the muon decay channel and a preliminary 
spectrum of events in elastic scattering as a function 
of the four-momentum transfer to the proton.

% ******************************************************************
% ******************************************************************
\subsection{\boldmath Diffractive $\bf Z$ Boson Production}

\begin{figure}
\vspace*{0.0cm}
\includegraphics[width=11cm]{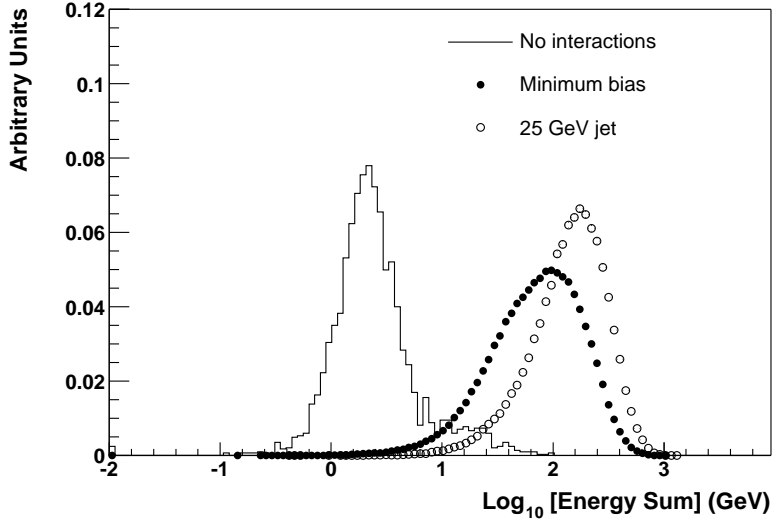}
\caption{
  \label{fig:Esum1} The logarithm of the reconstructed energy
       in the direction of the outgoing antiproton in the range
       $-5.3 < \eta < -2.6$.
       Compared are the distributions for events
       with no visible interactions (line) and
       events in which the proton and the anti-proton dissociate
       (full and open markers).
       The areas are normalised to unity.
       Rapidity gap candidates are selected by applying a cut at 10~GeV.} 
\end{figure}

D\O\  collected an event sample of about $21$~pb$^{-1}$
during Run I suitable for diffractive physics studies.
Nine diffractively produced $Z\rightarrow e^+e^-$
candidates were observed in this data sample~\cite{run1diff}. 
The current search for $Z$ bosons, produced in single diffraction, 
uses a dataset of $\approx 110$~pb$^{-1}$.
The analysis is based on the experimental reconstruction of a rapidity gap
close to the direction of the final-state (anti-) proton
and the reconstruction of the $Z$ via its decay into 
two muons with opposite charge.
Both muons were required to have $p_T > 15$~GeV.
One of the muons must be isolated in the calorimeter and in the 
central tracking detector.
Muons from cosmic rays were suppressed by the requirement that their
tracks stem from the vertex of the $p\bar{p}$ interaction
and that their acolinearity ($\Phi = \Delta \phi_{\mu \mu} +
\Delta \theta_{\mu \mu} - 2 \pi$) is 
larger than $0.05$~\cite{lyon,edwards04}.
The search of the rapidity gaps uses two scintillating detectors, 
one on each side of the interaction region, 
which cover the pseudorapidity range $2.7 < |\eta| < 4.4$. 
Calorimeter energies towards the directions of the
outgoing protons (in the range $2.6 < |\eta| < 5.3$) 
are summed separately on each side
with cell thresholds of $100$~MeV in the electromagnetic layers
and $200$~MeV in the fine hadronic layer.
The distribution of the logarithm of the energy sum is shown
in Fig.~\ref{fig:Esum1} for bunch crossings with no visible 
interactions.
Based on a random trigger 
these events are expected to have similar topologies as rapidity gap
events with no activity in the direction of the outgoing antiproton. 
Also shown is the energy distribution for a sample of minimum bias events.
A third event sample is defined by the requirement of (at least)
one jet with $p_T > 25$~GeV.
Events are excluded when the leading $p_T$ jet is in the region 
$|\eta| > 2.4$.
Both the minimum bias and the jet samples are dominated by 
non-diffractive events in which both protons
dissociate.
Events where the antiproton remains intact and 
events with antiproton dissociation can be separated by 
requiring the energy sum to be below 10~GeV.
The final selection of single diffractive $Z$ boson candidates
requires that on one side the scintillator is off and the energy sum is 
below 10~GeV while on the other side the scintillator is on
and the energy sum is larger than 10~GeV. 

\begin{figure}
  \includegraphics[scale=0.37]{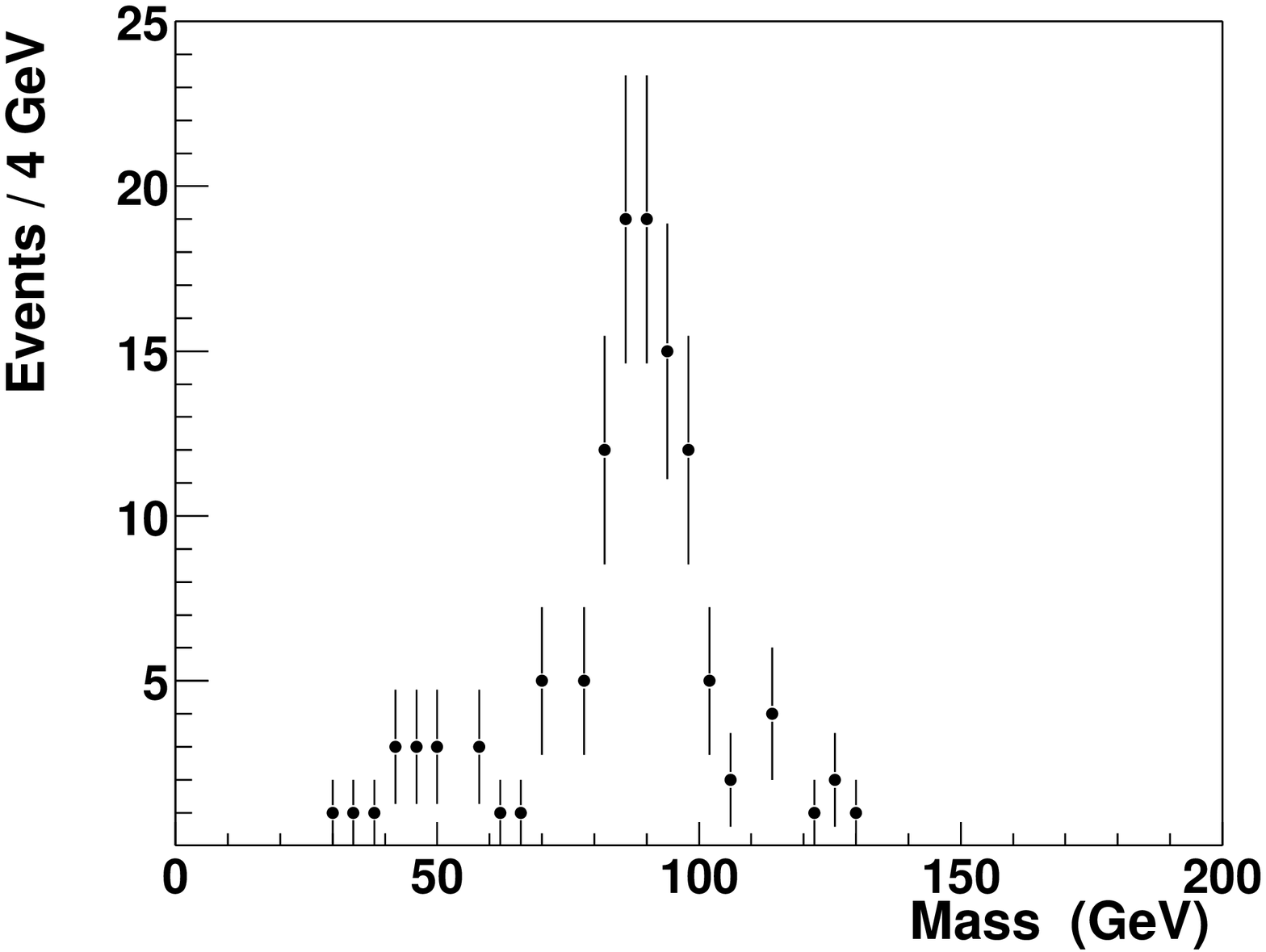}
  \includegraphics[scale=0.37]{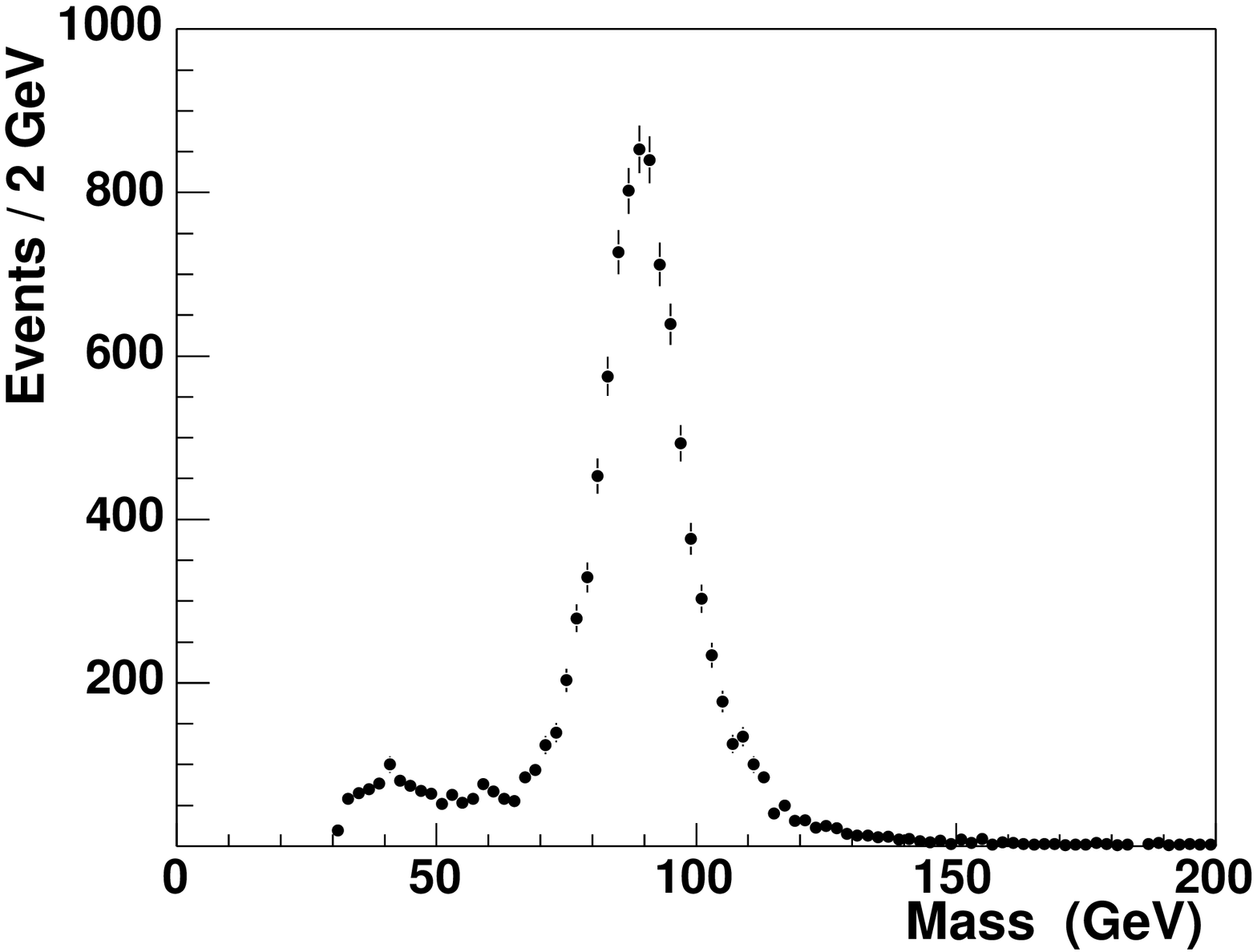}
   \caption{
 \label{fig:mumass} 
        The invariant di-muon invariant mass distribution for 
        $Z$ boson candidates with a single rapidity gap (left)
        and with no rapidity gap (right). 
        For details on the rapidity gap definition, please see the
        text.}
\end{figure}

The invariant di-muon mass distribution is shown
in Fig.~\ref{fig:mumass} for $Z$ boson candidates in events
with a rapidity gap (left) and in events that fail the 
rapidity gap selection (right).
The latter are strong candidates for non-diffractively produced 
$Z$ bosons.
Their mass distribution has a strong peak at the $Z$ mass
and small background.
The di-muon mass distribution for the candidates for diffractively 
produced $Z$ bosons has very similar shape.
Additional studies on the purity and the background for 
the rapidity gap selection are, however, still needed 
for quantitative statements and to establish the evidence
for a diffractive  $Z\rightarrow \mu^+ \mu^-$ signal.

% ******************************************************************
% ******************************************************************
\subsection{Elastic Scattering}

In the D\O\ experiment an intact outgoing (anti-) proton can be detected
in the Forward Proton Detector (FPD)~\cite{FPD}, which consists of
position detectors along the beam line in conjunction with accelerator 
magnets.
In total, the FPD has nine spectrometers, each comprising two 
scintillating fiber tracking detectors that can be moved to within a few 
millimeters of the beam. 
The kinematics of the outgoing (anti-) proton is described by
the four-momentum transfer squared, $t$ and the fractional longitudinal 
momentum loss $\xi$.
Both variables $|t|$ and $\xi$ can be computed from the
track four-vectors reconstructed in the FPD.

\begin{figure}
  \includegraphics[scale=0.53]{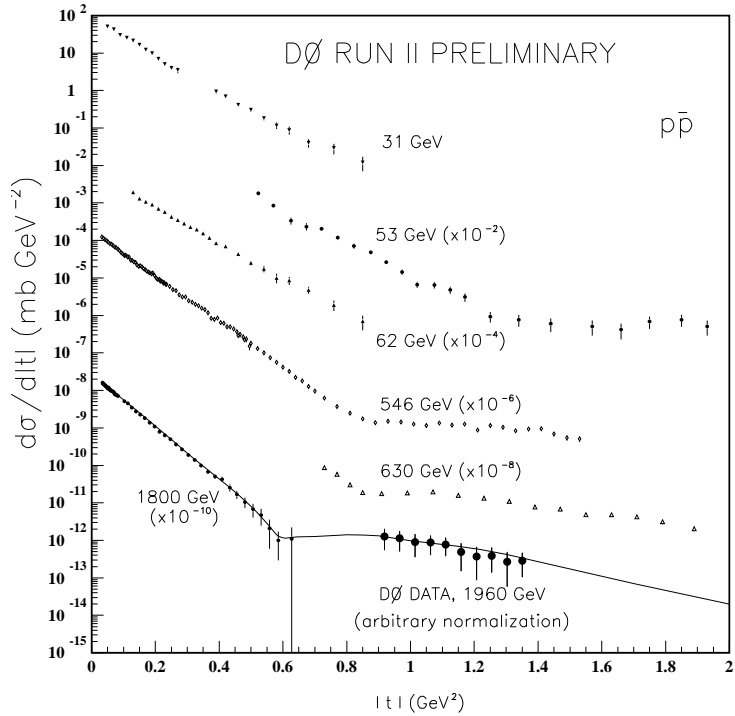}
\caption{
  \label{fig:dNdt} 
        The differential elastic $p{\bar p}$ scattering cross section 
        as a function
        of the four-momentum transfer squared $|t|$.
        Measurements from experiments at different center-of-mass 
        energies are shown (datasets at different energies are scaled 
        for visual separation).
        The preliminary D\O\ data are arbitrarily normalized
        for comparison of the slope with the E710 data at 
        $\sqrt{s}=1.8$\,TeV.
        Theoretical predictions from Block {\it et al.} are overlaid on 
        the data.
}
\end{figure}

We have measured the $|t|$ distribution in elastic scattering using the FPD.
Elastic events were triggered by two spectrometers which are above (below)
the beam on the side of the outgoing antiprotons (protons).
The spectrometer on the proton side was used to measure the
track position.
The trigger uses time-of-flight information, 
and vetoes events with hits in the scintillators 
covering $5.2 < |\eta| < 5.9$. 
The spectrometers can reconstruct protons in the kinematic range 
$0.8 \lesssim |t| \lesssim 4.0$~GeV$^2$, 
depending on detector position and accelerator conditions. 
The preliminary results are presented in the range
$0.92 < |t| < 1.34$~GeV$^2$.

A Monte Carlo event generator was used 
to propagate the outgoing (anti-\/) protons in elastic
scattering through the magnetic field of the Tevatron
and to simulate the response of the FPD based on the expected detector 
resolution~\cite{backgroundcuts}. 
The geometric acceptance of the FPD was modeled by a similar
event generator~\cite{acceptance} as a function $|t|$.
Remaining backgrounds from beam halo~\cite{halo} and from 
single diffraction are subtracted separately.

The preliminary $dN/d|t|$ distribution for elastic $p{\bar p}$ 
scattering is shown in Fig.~\ref{fig:dNdt}.
Measurements of the differential cross section $d\sigma/d|t|$
at different center-of-mass energies from the ISR, UA4, and 
E710 experiments are also shown~\cite{OtherData}.
The normalization of the D\O\ data is arbitrary, but it is
visible that the shape of the D\O\ data in the range 
$0.8 < |t| < 1.4$\,GeV is very 
different from the shape of the E710 data measured at 
$\sqrt{s}=1.8$\,TeV and at $|t| < 0.65$\,GeV.
A theoretical prediction by Block {\it et al.}~\cite{Block}
is compared to the data at highest center-of-mass energies.
The model predicts a transition between the
different $|t|$ regions and give a good description of the data
over the whole $|t|$ range.

%%%%%%%%%%%%%%%%%%%%%%%%%%%%%%%%%%%%%%%%%%%%%%%%
%% BACKMATTER
%%%%%%%%%%%%%%%%%%%%%%%%%%%%%%%%%%%%%%%%%%%%%%%%

%% If the bibliography is produced without BibTeX comment out the

\bibliographystyle{aipproc}   % if natbib is available

\bibliography{sample}

\IfFileExists{\jobname.bbl}{}
 {\typeout{}
  \typeout{******************************************}
  \typeout{** Please run "bibtex \jobname" to obtain}
  \typeout{** the bibliography and then re-run LaTeX}
  \typeout{** twice to fix the references!}
  \typeout{******************************************}
  \typeout{}
 }

\end{document}